\documentclass[prb,twocolumn,showpacs,notitlepage]{revtex4}
\usepackage[cp1251]{inputenc}
\usepackage{amsmath}
\usepackage{amssymb}
\usepackage{hyperref}
\usepackage{natbib}
\usepackage{bm}
\usepackage{epsfig}
\begin{document}

 \renewcommand {\Im}{\mathop\mathrm{Im}\nolimits}
  \renewcommand {\Re}{\mathop\mathrm{Re}\nolimits}
    \newcommand {\Tr}{\mathop\mathrm{Tr}\nolimits}
\newcommand {\rmi}{{\rm i}}
\newcommand {\rmd}{{\rm d}}
\newcommand {\sign}{\mathop{\mathrm{sign}}\nolimits}
\newcommand {\e}{{\rm e}}
\renewcommand {\phi}{\varphi}
\renewcommand {\epsilon}{\varepsilon}
\newcommand {\eps}{\varepsilon}
\newcommand{\nix}[1]{}
\title{Collective F\"orster energy transfer modified by the planar metallic mirror}
 \author{Alexander N. Poddubny}
 \affiliation{Ioffe Institute, St Petersburg 194021, Russia}
  \affiliation{ITMO University, St Petersburg 199034, Russia}
 \email{poddubny@coherent.ioffe.ru}
\begin{abstract}
We present a theory of the F\"orster energy transfer between the arrays of donor and acceptor molecules lying on the planar 
metallic mirror. We reveal strong modification of the effective transfer rate by the mirror in the incoherent pumping regime. The rate can be either suppressed or enhanced depending on the relative positions between acceptor and donor arrays. The strong modification of the transfer rate is a collective effect, mediated by 
the light-induced coupling between the donors; it is absent in the single donor model.
\end{abstract}
\date{\today}
\pacs{78.67.Pt,78.67.-n,33.50.-j,42.50.Nn}
%78.67.-n	Optical properties of low-dimensional, mesoscopic, and nanoscale materials and structures

%78.67.Pt 	Multilayers; superlattices; photonic structures; metamaterials (see also 81.05.Xj, Metamaterials for chiral, bianisotropic and other complex media)
%33.50.-j	Fluorescence and phosphorescence; radiationless transitions, quenching (intersystem crossing, internal conversion) (for energy transfer, see also section 34; for biophysical applications, see 87.64.kv)

%42.50.Nn	Quantum optical phenomena in absorbing, amplifying, dispersive and conducting media; cooperative phenomena in quantum optical systems

\maketitle

%%%%%%%%%%%%%%%%%%%%%%%%%%%%%%%%%%%%%
\section{Introduction}\label{sec:intro}
F\"orster energy transfer  processes are now actively studied in various fields that bridge  physics, biology and medicine. The energy is transferred from the excited (donor) system to the acceptor one via the electromagnetic interaction. The donor and acceptor systems may be realized as  quantum dots and quantum wells,\cite{Agranovich2011,Andreakou,Rindermann} biological molecules,\cite{Fruhwirth,Lopez2014} defects in semiconductor.\cite{DaldossoPRB2009,ProkofievTransferPRB2008}  Typically, the range of the F\"orster interaction is on the order of several nm.\cite{AgranovichGalanin} 

 One can try to control the efficiency of the transfer by embedding the donors and acceptors into the structured electromagnetic environment. Contradicting experimental reports of the  enhanced\cite{Andrew2000}, modulated \cite{Nakamura2005}, independent \cite{rabouw2014,blum2012} and suppressed\cite{We2014} transfer are available in literature. In the particular  case of Ref.~\onlinecite{We2014} the transfer has been studied for the dye molecules embedded in the thin polymer film. The speedup of the donor emission decay in the presence of the acceptors has been attributed to the energy transfer rate. The transfer rate has been suppressed when the film was put on top of the metallic mirror. 
As has been pointed out in Ref.~\onlinecite{blum2012}, for the planar mirror this effect cannot be explained by the existing well-developed theory of energy transfer between individual independent pairs  of donors and acceptors. \cite{Andrews1994,Dung2002,Andrews2004,Shahbazyan2011} It has been suggested in Ref.~\onlinecite{We2014}, that the observed transfer suppression can be due to the collective effects, since the concentration of the donor dye molecules was quite large and the intermolecular distance was on the order of several nm. However, any further theoretical explanation was missing in Ref.~\onlinecite{We2014}. 
Developing a theory of collective energy transfer from donors to acceptors in the presence of a planar mirror is the main goal of the current study. Recently, similar theory has been put forward  by Pustovit, Urbas and Shahbazyan in Ref.~\onlinecite{Shahbazyan2013} for  molecules surrounding the spherical metallic particles. While the general approach is similar, the electromagnetic modes for the spherical geometry are quite different, so the results of Ref.~\onlinecite{Shahbazyan2013} cannot be applied to our system directly. Hence, there is a growing demand from both theory and experiment to elaborate on the energy transfer in planar plasmonic systems with large molecule concentration.

In the rest of the paper we present the model (Sec.~\ref{sec:Model}) for the collective transfer and show, that the  transfer rate can be strongly modified by the mirror (Sec.~\ref{sec:Results}). 
%%%%%%%%%%%%%%%%%%%%%%%%%%%%%%%%%%%%%%%%%%%%%%%%%%%
\begin{figure}
\centering\includegraphics[width=0.4\textwidth]{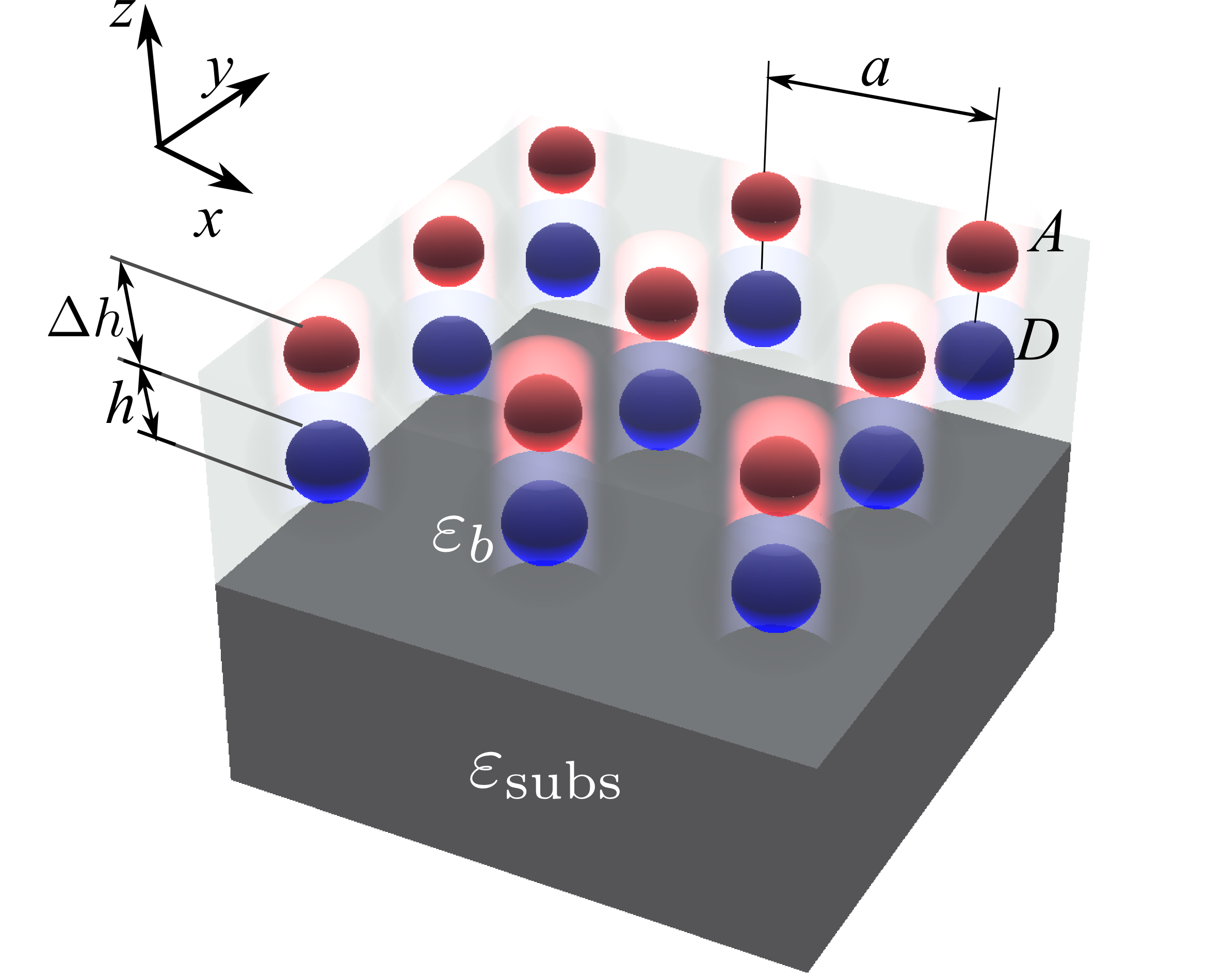}
\caption{Schematic illustration of the studied system. Arrays of donor (D) and acceptor (A) molecules above the substrate are indicated.}\label{fig:geometry}
\end{figure}
%%%%%%%%%%%%%%%%%%%%%%%%%%%%%%%%%%%%%%%%%%%%%%%%%%%
\section{Model}\label{sec:Model}
%%%%%%%%%%%%%%%%%%%%%%%%%%%%%%%%%%%%%%%%%%%%%%%%%%%
The system under consideration is schematically depicted in Fig.~\ref{fig:geometry}. It consists of the  metallic substrate with the permittivity $\eps_{\rm subs}$ bounded from top with the semi-infinite dielectric matrix with the permittivity $\eps_{b}$.  Arrays of donor and acceptor molecules form square lattices with the period $a$, that are parallel to the substrate. The donor array is placed at the height $h$ from the substrate, the acceptor array lies in the plane $z=h+\delta h$ at the vertical distance $\delta h$ from the acceptor array. The $x$ and $y$ axes are parallel to the interface plane $z=0$.

We use the semiclassical coupled dipole approximation with the random source terms, similarly to the approach in Refs.~\onlinecite{JETP2009,Shahbazyan2013}:
\begin{align}
\frac{1}{\alpha_{D}}\bm p_{D,j}&=\bm f_{D,j}+\sum\limits_{j'\ne j}G(\bm r_{D,j}-\bm r_{D,j'})\bm p_{D,j'}\:,\label{eq:syst1}\\
\frac{1}{\alpha_{A}}\bm p_{A,j}&=\sum\limits_{j'}G(\bm r_{A,j}-\bm r_{D,j'})\bm p_{D,j'}\:.\label{eq:syst2}\\
\end{align}
Here $\bm p_{D,j}$ ($\bm p_{A,j}$) are the dipole momenta of the donors (acceptors) located at the points $\bm r_{D,j}$ ($\bm r_{A,j}$) and having the polarizabilities $\alpha_{D}$ ($\alpha_{A}$), $G(\bm r,\bm r')$ is the tensor Green function
\begin{equation}
\nabla\times\nabla\times G(\bm r,\bm r')=\left(\frac{\omega}{c}\right)^{2}\eps(\bm r)G(\bm r,\bm r')+4\pi\left(\frac{\omega}{c}\right)^{2} \hat 1\delta(\bm r-\bm r')\:,
\end{equation}
accounting for the presence of the metallic mirror,\cite{Tomas1995} and $\bm f_{D,j}$ are the random source terms.
We assume incoherent stationary pumping of the donor array, in this case the realization-averaged values of $\bm f_{D,j}$ are equal to zero. The pumping to different molecules can be considered independent and is characterized by the following correlator:
\begin{equation}
\langle f_{D,j,\mu}f_{D,j',\nu}\rangle=\frac{S}{2\pi}\delta_{\mu\nu}\delta_{jj'}\delta(\omega-\omega')\:,\label{eq:S}
\end{equation}
here $S$ is the effective rate of the generation of excitons in donors~\cite{JETP2009}, $\mu,\nu=x,y,z$ and the brackets denote the averaging over the realizations of the random sources.
The polarizabilities  have the resonant form,  
\begin{equation}
\alpha_{M}=\frac{d_{M}^{2}}{\hbar[\omega_{M}-\omega-\rmi (\Gamma_{M}+\Gamma_{0,M})]},\quad M=D,A\:,
\end{equation}
where $d_{M}$ are the matrix elements of the dipole moment for donors and acceptors, $\omega_{M}$ are the resonant frequencies and $\Gamma$ are the phenomenological nonradiative decay rates. The radiative decay rate $\Gamma_{0}$ is then described by the standard expression\cite{Novotny2006}
\begin{equation}
\Gamma_{0,M}=\frac{2\omega_{M}^{3}d_{M}^{2}\sqrt{\eps_{b}}}{3\hbar c^{3}}\:.
\end{equation}
Our goal is to determine the stationary population of acceptors induced as a result of the pumping of the donors and the energy transfer from donors to acceptors. 
Importantly, in Eq.~\eqref{eq:syst1} we include the electromagnetic  coupling between the donor molecules (second term in the right hand side). This coupling  leads to the formation of the collective states, eigenmodes of Eq.~\eqref{eq:syst1}.  Importantly, the collective effects are manifested even for incoherent pumping. The only necessary conditions are the weak inhomogeneous broadening in the donor array and long enough lifetime of the donor states.  In the linear-in-pumping regime the pumping determines only the population of the different eigenmodes, but not their spatial structure. In order to simplify the model  we neglect the electromagnetic coupling between acceptors in Eq.~\eqref{eq:syst2}.  The reason is that the energy is transferred to the excited states of the acceptors. The nonradiative decay rate $\Gamma_{A}$ of the excited acceptor state is assumed enough fast to quench both the acceptor-acceptor coupling and possible transfer of the energy back to the donors. 

We will first calculate the population of donors by solving Eq.~\eqref{eq:syst1}. This can be accomplished by expanding the donor dipole momenta over the Bloch modes, characterized by the in-plane Bloch wave vector $\bm k$:
\begin{multline}
\bm p_{D,j}=\sum_{\bm k}\e^{\rmi \bm k\bm r_{D,k}}\bm p_{D,\bm k}\\\equiv
\frac{a^{2}}{(2\pi)^{2}}\iint\limits_{|k_{x,y}|<\pi/a}\rmd k_{x}\rmd k_{y}\e^{\rmi \bm k\bm r_{D,k}}\bm p_{D,\bm k}\:.\label{eq:Fourier}
\end{multline}
Substituting Eq.~\eqref{eq:Fourier} into Eq.~\eqref{eq:syst1} we obtain a system of independent equations for the Bloch amplitudes $\bm p_{D,\bm k}$. Solving them and substituting the result back to Eq.~\eqref{eq:Fourier} we obtain 
\begin{equation}
\bm p_{D,j}=\sum\limits_{\bm k} \sum\limits_{j'}\e^{\rmi \bm k(\bm r_{D,j}-\bm r_{D,j'})}\tilde  \alpha_{D,\bm k}\bm f_{D,j'}\:.\label{eq:pDj}
\end{equation}
where
\begin{equation}
\tilde  \alpha_{D,\bm k}=
\frac{\alpha_{D}}{1-\alpha_{D}C_{D,\bm k}}\:,\label{eq:aDk}
\end{equation}
and
\begin{equation}
 C_{D,\bm k}=\sum\limits_{j\ne 0}\e^{-\rmi\bm k\bm r_{D,j}}G(-\bm r_{D,j})\label{eq:C}
\end{equation}
is the interaction constant for the planar donor array.\cite{Belov2005} Eq.~\eqref{eq:pDj} fully accounts for the hybridization between the donor array and the electromagnetic modes of the metallic substrate, including the plasmonic modes. This is encoded in the polarizabilities of the Bloch modes $\tilde\alpha_{D,\bm k}$, which have the form of the bare donor polarizability $\alpha_{D}$ renormalized by the interaction with the mirror. The interaction constant $C_{D,\bm k}$ can be evaluated either by Floquer summation \cite{Belov2005} or by an Ewald-type summation \cite{Kambe1967}, here we resort to the latter procedure.

The average occupation number of the donor molecules can be introduced as the averaged squared amplitude of the donor dipole momentum divided by the dipole momentum matrix element:
\begin{multline}
N_{D}\equiv\frac1{|d_{D}|^{2}} \langle |\bm p_{D}(t)|^{2}\rangle\\=
\frac1{|d_{D}|^{2}} \iint \frac{\rm d\omega\rm d\omega'}{(2\pi)^{2}}\e^{\rmi (\omega-\omega')t} \left\langle\bm p^{*}_{D,j}(\omega)\cdot\bm p_{D,j}(\omega')\right\rangle\:.\label{eq:ND0}
\end{multline}
Substituting Eq.~\eqref{eq:pDj} into Eq.~\eqref{eq:ND0} and using the random source correlation function Eq.~\eqref{eq:S} we obtain
\begin{equation}
N_{D}=S\int \frac{\rmd\omega}{2\pi} \sum\limits_{\bm k} \Tr\tilde  \alpha^{\vphantom{\dag}}_{D,\bm k}  \tilde\alpha^{\dag}_{D,\bm k} \:.\label{eq:ND1}
\end{equation}
Now we will assume weak coupling between the donor array and the mirror. In this approximation the interaction with the mirror leads only to the modification of the donor lifetime. The complex polarizabilities $\tilde\alpha_{D,\bm k}$ then have the form
\begin{equation}
\tilde\alpha_{D,\bm k}=\frac{d_{D}^{2}}{\hbar[\omega_{M}-\omega-\rmi (\Gamma_{D}+F_{\bm k}\Gamma_{0,D})]}\:,
\end{equation}
where 
\begin{equation}
F_{\bm k}=1+\frac{3}{2(\omega_{D}/c)^{3}\sqrt{\eps_{b}}}\Im C_{\bm k}(\omega_{D}),\label{eq:F}
\end{equation}
is the tensor analogue of the Purcell factor, describing the lifetime modification.
The matrix $F_{\bm k}$ can be diagonalized,
\begin{equation}
F_{\bm k,\mu\nu}=\sum\limits_{\lambda=1}^{3}V_{\bm k,\mu\lambda}f_{\bm k,\lambda}[V_{\bm k}^{-1}]_{\lambda\nu},\quad \mu,\nu=x,y,z\:,\label{eq:F}
\end{equation}
where $f_{\bm k, \gamma}$ are the Purcell factors characterizing the decay of the Bloch modes with different polarizations. There exist three possible polarizations for each value of $\bm k$. Note, that the matrix $V_{\bm k}$ in Eq.~\eqref{eq:F} is in general case neither Hermitian nor symmetric.
After the frequency integration in Eq.~\eqref{eq:ND1} is carried out, the result for the donor population assumes the form
\begin{equation}
N_{D}= S\sum\limits_{\bm k,\lambda}\tau_{\bm k,\lambda}\label{eq:ND}
\end{equation}
where
\begin{equation}
\tau_{\bm k,\lambda}=\frac1{2(\Gamma_{D}+f_{\bm k,\lambda}V^{*}_{\bm k,\lambda\mu}[V_{\bm k}^{-1}]_{\mu\lambda})}\label{eq:tDk}
\end{equation}
is the effective lifetime of the Bloch mode and the summation over the dummy index $\mu$ is assumed.
Eq.~\eqref{eq:ND} has a clear physical meaning: the total population of the donors is a sum over the populations of different eigenmodes. Each eigenmode population is given by the product of the pumping rate and the lifetime. Since donors are pumped in an uncorrelated way, the pumping constant is independent of the Bloch vector $\bm k$. However,  the lifetimes $\tau_{\bm k,\lambda}$ can differ for different modes. 

Now we proceed to the calculation of the acceptor population. To this end we substitute Eq.~\eqref{eq:pDj} into Eq.~\eqref{eq:syst2} and calculate the correlation function
\begin{multline}
N_{A}\equiv\frac1{|d_{A}|^{2}} \langle |\bm p_{A}(t)|^{2}\rangle\\=
\frac1{|d_{A}|^{2}}\iint \frac{\rm d\omega\rm d\omega'}{(2\pi)^{2}}\e^{\rmi (\omega-\omega')t} \left\langle \bm p^{*}_{A,j}(\omega)\cdot\bm p_{A,j}(\omega')\right\rangle\:.\label{eq:NA0}
\end{multline}
This yields (cf. with Eq.~\eqref{eq:ND1} and see also Ref.~\onlinecite{Vergeer2005})
\begin{multline}
N_{A}=S\int \frac{\rmd\omega}{2\pi} |\alpha_{A}|^{2}\sum\limits_{\bm k} \Tr
G_{DA,\bm k}\ \alpha^{\vphantom{\dag}}_{D,\bm k} \alpha^{{\dag}}_{D,\bm k}G^{\dag}_{DA,\bm k}
\label{eq:NA1}
\end{multline}
where
\begin{equation}
G_{DA,\bm k}=\sum\limits_{j}\e^{\rmi\bm k\bm r_{A,j}}G(\bm r_{A,j}-\bm r_{D,j'})\:
\end{equation}
is the Green function Fourier component describing the donor-acceptor coupling.
In the chosen approximation the lifetime of the (excited) acceptor states is determined solely by energy relaxation processes,
$\Gamma_{A}\gg\Gamma_{0,A},\Gamma_{0,D},\Gamma_{D},1/\tau_{\bm k,\lambda}$.
Hence, we can introduce the effective energy transfer rate $1/\tau_{ET}$ from the following kinetic equation for the acceptors:
\begin{equation}
\frac{N_{A}}{\tau_{A}}=\frac{N_{D}}{\tau_{ET}}\:.\label{eq:tEt}
\end{equation}
The left-hand side of this equation describes the (nonradiative) decay of acceptors with the lifetime $\tau_{A}=1/(2\Gamma_{A})$, the right-hand-side describes the transfer of energy from donors with the population $N_{D}$.
Integrating over frequency in Eq.~\eqref{eq:NA1}, using the donor population $N_{D}$ from Eq.~\eqref{eq:ND} and the transfer rate definition Eq.~\eqref{eq:tEt} we obtain
\begin{equation}
\frac1{\tau_{ET}}=\frac{2\pi}{\hbar}\frac1{\pi}\frac{\Gamma_{A}}{(\omega_{D}-\omega_{A})^{2}+\Gamma_{A}^{2}}
%:
|d_{A}|^{2}|d_{D}|^{2}\mathcal G^{2}\:,\label{eq:tET1}
\end{equation}
where the effective constant of the transfer rate reads
\begin{equation}
\mathcal G^{2}=\frac{W}{T}\equiv\frac{1}{\sum\limits_{\bm k,\lambda}\tau_{\bm k,\lambda}}
\sum\limits_{\bm k,\lambda}
\tau_{\bm k,\lambda}
|\bm G_{DA,\bm k,\lambda}|^{2}\label{eq:G2}
\end{equation}
with
\begin{equation}
\bm G_{DA,\bm k,\lambda}=\bm e_{\mu }[G_{DA,\bm k}]_{\mu\nu}V_{\bm k,\nu\lambda}\:.\label{eq:GDAkl}
\end{equation}
Equations~\eqref{eq:tET1}--\eqref{eq:GDAkl} for the effective energy transfer rate constitute the central result of this study.
Their structure is quite clear. The first resonant factor in
Eq.~\eqref{eq:tET1} depends on the spectral detuning between donor and acceptor molecules. The second factor $\mathcal G^{2}$ describes the electromagnetic coupling between the molecules arrays. If the collective coupling between the donors is neglected, the factor reduces to 
\begin{equation}
\mathcal G_{0}^{2}=\sum\limits_{j'}\Tr [G^{\dag}(\bm r_{A,j}-\bm r_{D,j'})G(\bm r_{A,j}-\bm r_{D,j'})]\:, \label{eq:GDAkl0}
\end{equation}
i.e. it is described by a sum of individual transfer events from different donors. In this case the transfer rate assumes the same form as in Refs.~\onlinecite{Andrews1994,Dung2002,Andrews2004}. In the general case, however, the transfer constant $\mathcal G^{2}$ is given by a sum of the contributions from Bloch eigenmodes of the donor ensemble. Since the donors are pumped incoherently and uncorrelated, the contributions 
are independent. Each term in the sum in Eq.~\eqref{eq:tET1} is proportional to the lifetime of the corresponding mode $\tau_{\bm k,\lambda}$, that determines the mode population.

%%%%%%%%%%%%%%%%%%%%%%%%%%%%%%%%%%%%%%%%%%%%%%%%%%%
\section{Results and discussion}\label{sec:Results}
%%%%%%%%%%%%%%%%%%%%%%%%%%%%%%%%%%%%%%%%%%%%%%%%%%%
In this section we present detailed analysis of the effect of the metallic mirror on the transfer rate 
Eq.~\eqref{eq:G2}. 
%%%%%%%%%%%%%%%%%%%%%%%%%%%%%%%%%%%%%%%%%%%%%%%%%%%
\begin{figure}[t!]
\centering\includegraphics[width=0.45\textwidth]{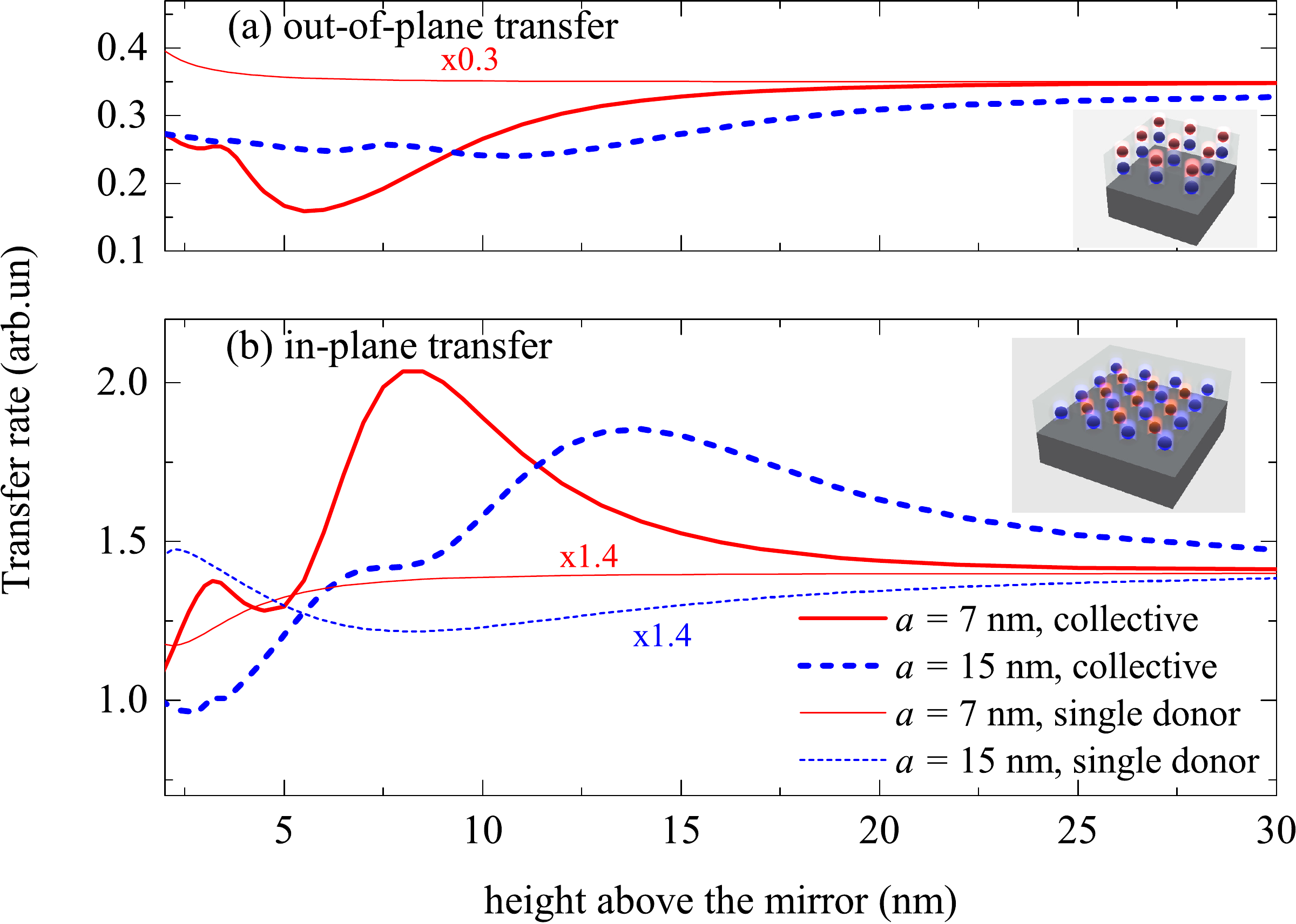}
\caption{(Color online) Energy transfer rate as function of the height of the donors above the mirror $h$ for (a) acceptors located above donors and (b) acceptors located in the same horizontal plane as donors.
Thick solid/red  and dashed/blue curves correspond to the transfer rates calculated according to Eq.~\eqref{eq:G2} with the lattice constants $a=7~$nm and $a=15~$nm, respectively.  The curves are normalized to the transfer rates between single donor and single acceptor at the corresponding distance without the mirror. Thin curves show the transfer rates between single donor and single acceptor, i.e. calculated from Eq.~\eqref{eq:GDAkl0} neglecting the collective effects.
The calculation parameters are as follows:  $\lambda=2\pi c/\omega_{D}=500$~nm, $\eps_{b}=2$, $\Gamma_{D}=0.1\Gamma_{0,D}$, $\eps_{\rm subs}=-9.77+0.31\rmi$ (corresponds to Ag in Ref.~\onlinecite{JohnsonChristy}), $\bm r_{A,j}-\bm r_{D,j}=\Delta h \hat{\bm z}$, $\Delta h=3.5$~nm (a) and  $\bm r_{A,j}-\bm r_{D,j}=a/4(\hat{\bm x}+\hat{\bm y})$ (b). The insets schematically illustrate the geometry of the problem.
}\label{fig:2transfer}
\end{figure}
Figure~\ref{fig:2transfer} shows the dependence of the transfer rate on the height of the donor and acceptor molecules above the mirror. The lattice constant of the donor and acceptor arrays and their relative positions with respect to each other remain fixed. Two panels of Fig.~\ref{fig:2transfer}  correspond to different geometries: (a) out-of-plane transfer, when acceptors are positioned above the donors at the height $\delta h$ and (b) in-plane transfer, when acceptors are positioned in the same horizontal plane as the donors so that each acceptor lies in the center of the square unit cell of the donor lattice.
Thick lines are calculated according to Eq.~\eqref{eq:GDAkl} and fully account for the hybridization between the donors while thin lines are calculated in the independent donor approximation Eq.~\eqref{eq:GDAkl0} by retaining only one term for one  donor, that is closest to the acceptor.
Importantly, the transfer rates in the collective model and the independent donor model are different from each other
even in the absence of the mirror (or for $h\gg  a$). The reason is that  the local field of the donor array at the acceptor is different from that of the single donor even in vacuum. In order to compare the two models visually, the curves for the single-donor model have been scaled by the factors $0.3$ and $1.4$ in Fig.~\ref{fig:2transfer}a and Fig.~\ref{fig:2transfer}b, respectively. 
The main result of this work, shown in Fig.~\ref{fig:2transfer}, is as follows: the collective transfer rate is strongly sensitive to the distance from the mirror (thick lines) while the  independent-donor one is not. When the acceptors are positioned above the donors, the transfer is strongly suppressed by the mirror, while for acceptors located between the donors the transfer is enhanced. This effect cannot be described in the independent transfer model (thin curves), which predicts almost vanishing dependence of the transfer rate on the distance up to the distance $h\sim 3~$nm, in agreement with Refs.~\onlinecite{blum2012,We2014}. The transfer modification is most pronounced when the height of the donor array above the mirror is on the order of the array period $a$ (cf. solid and dotted curves for $a=7$~nm and $a=15$~nm). As the period increases, the effect is suppressed.
Thus, collective transfer model predicts a strong modification of the transfer rate near the metallic mirror, in stark contrast to the independent transfer model. Now we  proceed to the explanation of these findings.

%%%%%%%%%%%%%%%%%%%%%%%%%%%%%%%%%%%%%%%%%%%%%%%%%%%
\begin{figure}
\centering\includegraphics[width=0.4\textwidth]{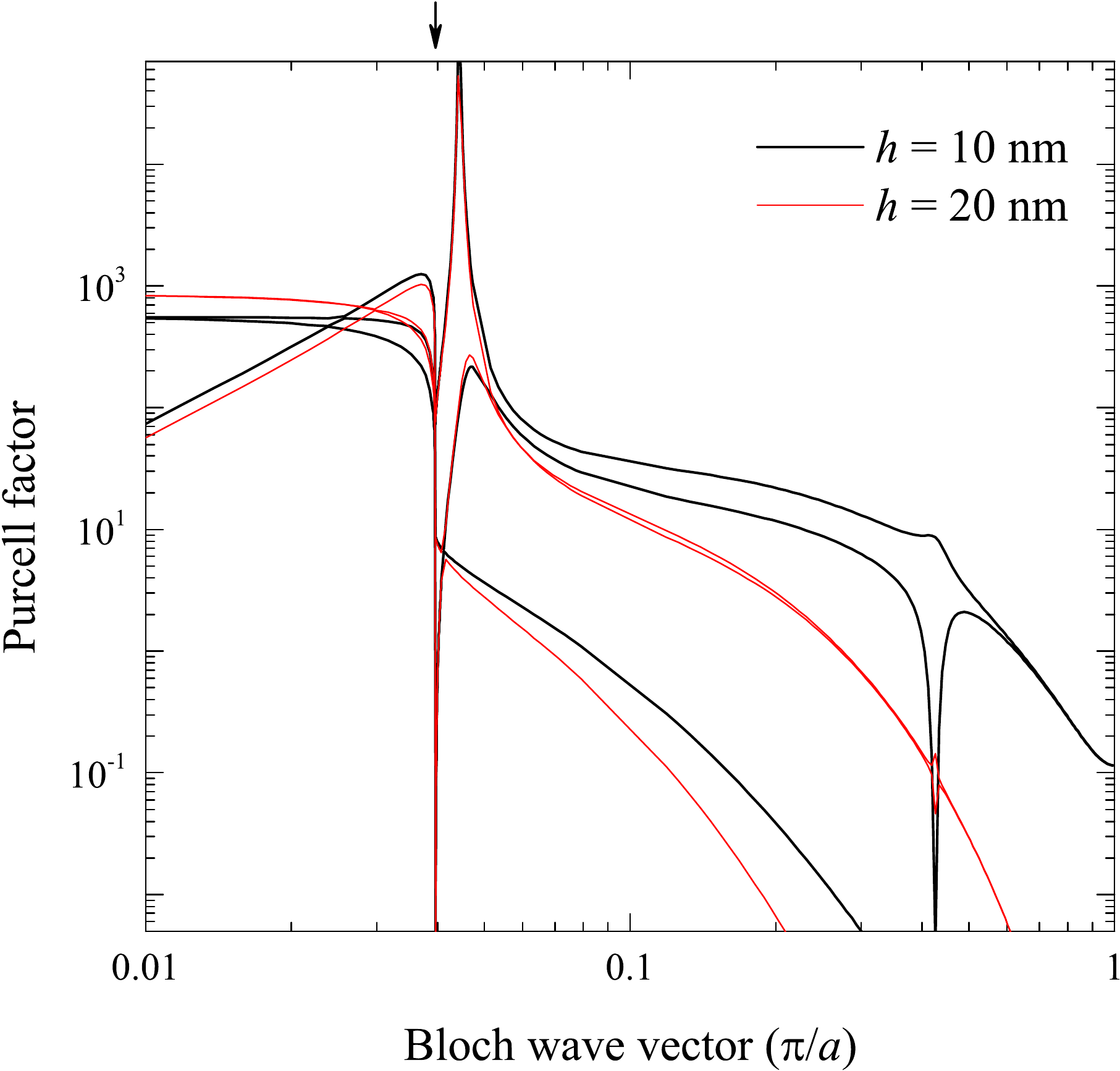}
\caption{
Purcell factors for the Bloch modes of the donor array with different polarizations as functions of the Bloch wave vector along the $\Gamma$--X direction. Thick/black and thin/red curves correspond to $h=10$~nm and $h=20$~nm, respectively.
Calculated for $a=7$~nm and the same other parameters as in Fig.~\ref{fig:2transfer}. The arrow indicates the value of $k$ equal to the light cone edge $\omega_{D}\sqrt{\eps_{b}}/c$.
}\label{fig:3times}
\end{figure}
%%%%%%%%%%%%%%%%%%%%%%%%%%%%%%%%%%%%%%%%%%%%%%%%%%%
\begin{figure}[b]
\centering\includegraphics[width=0.3\textwidth]{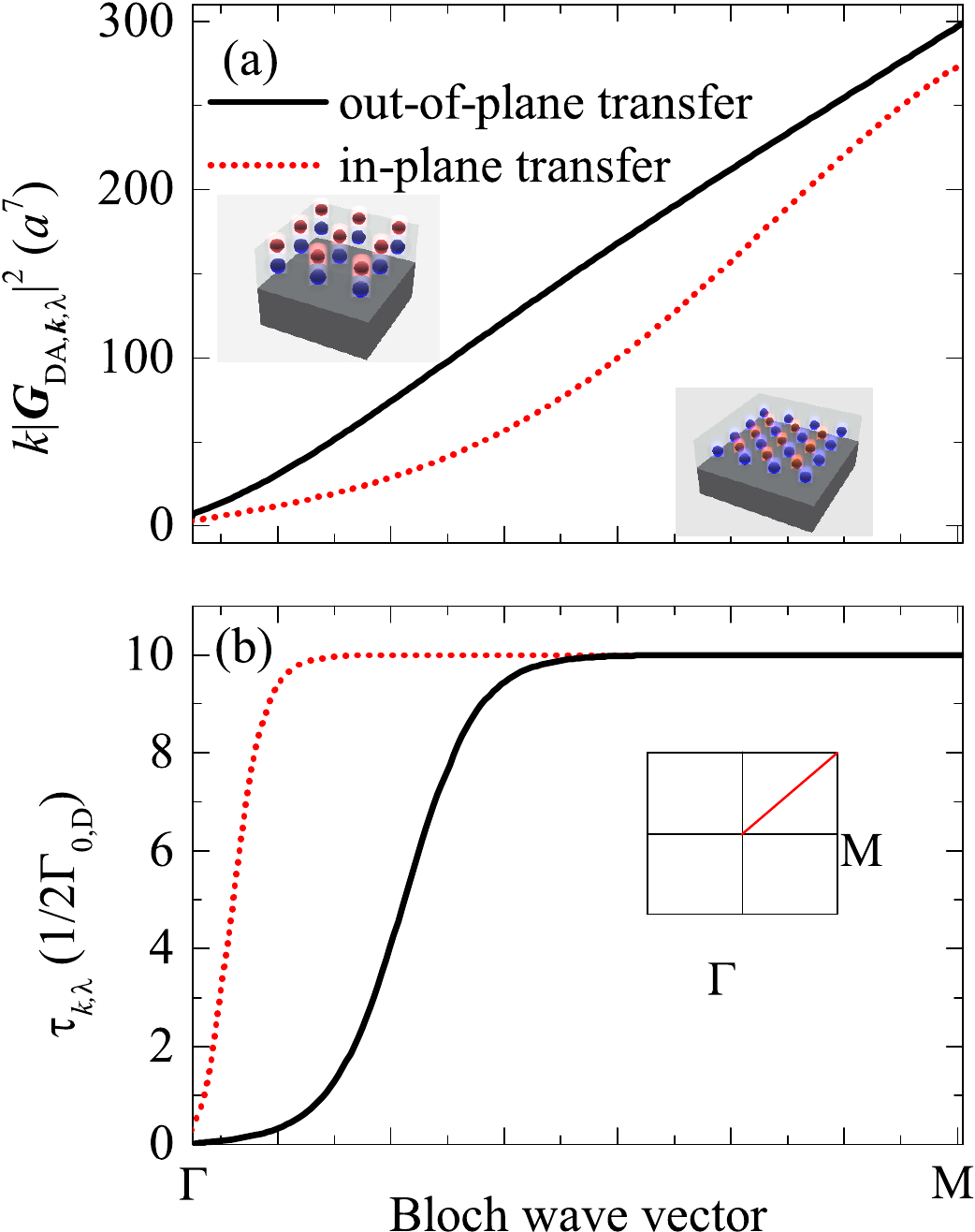}
\caption{
Green function Eq.~\eqref{eq:GDAkl} characterizing the transfer strength (a) and  lifetimes of the Bloch modes of the donor array (b) as functions of the Bloch wave vector in the $\Gamma$--M direction [inset in panel (b)]. Black/solid and red/dotted curves correspond to the out-of-plane energy transfer and in-plane transfer, as indicated by the insets in panel (a). For each geometry only the mode with one polarization, providing the largest contribution to the energy transfer, is shown.
Other calculation parameters as the same as in Fig.~\ref{fig:2transfer}.
}\label{fig:4Green}
\end{figure}

Our analysis indicates, that the Green functions, entering Eq.~\eqref{eq:GDAkl}, are not very sensitive to the distance from the mirror. Hence, the strong transfer modification, observed in Fig.~\ref{fig:2transfer} is solely due to the dependence of the lifetimes of the Bloch eigenmodes $\tau_{\bm k,\lambda}$ on the distance. The Purcell factors $f_{\bm k,\lambda}$, characterizing the lifetime modification by the mirror, are presented in Fig.~\ref{fig:3times}. Thick and thin curves were calculated as functions of the Bloch wave vector for two distances, $h=10$~nm and $h=20$~nm, respectively. For each height we show three curves, corresponding to different eigenmode polarizations. The calculation reveals strong dependence of the Purcell factors on the wave vector, distance and polarization. For given distance one can distinguish between several regions in Fig.~\ref{fig:3times}. The first region is inside the light cone ($k<\omega/c$). In this case the Purcell factor is much larger than unity even at the large distances from the mirror (or without the mirror). This can be understood as a semiclassical counterpart of the Dicke superradiance,\cite{Khitrova2007nat} when the donors, that are separated by the distances smaller than the light wave length, emit collectively. The radiative rate is larger by the factor  on the order of $1/(\omega a/c )^{2}$ than that of the individual donor in vacuum.\cite{Kavokin1991} Second, one can see the  peak in Fig.~\ref{fig:3times} at $k\sim 0.04\pi/a$, that describes strong enhancement of the donor decay due to the plasmons. The peak is present only for one of the curves, corresponding to the transverse magnetic  (TM, or $p$) polarization. Third, there exists a region of larger Bloch wave vectors $k\gtrsim 0.1 \pi/a$. This range corresponds to the evanescent modes, with the electric field exponentially decaying from the donor plane and lying outside the light cone. In the absence of the mirror these waves would be completely evanescent and they would have zero radiative decay rate, $f_{\bm k}=0$. Due to the presence of the mirror, the modes acquire finite lifetime due to the absorption  in the substrate, $f_{\bm k}\propto \Im\eps_{\rm subs}\e^{-2kh}$\:.\cite{Agranovich2011} The weakest values of the Purcell factor in this third region correspond to the donor modes with the polarization close to transverse electric  (TE, or $s$). Although  the Purcell factor for the evanescent modes is much less than unity, it is strongly enhanced when the array approaches the mirror  and the coupling to the mirror increases (cf. black and red curves). We will now explain how this very effect  leads to the modification of the transfer rate Eq.~\eqref{eq:G2} by the mirror. 

Figure~\ref{fig:4Green} shows the dependence of the Green function components Eq.~\eqref{eq:GDAkl} and the lifetimes of donor modes on the Bloch wave vector for out-of-plane (black solid curves) and in-plane transfer (red dotted curves).  For each geometry we have plotted the Green function component only for one polarization, that is largest by the absolute value and provides the dominant contribution to the transfer rate. The Green function values in panel (a) have been additionally multiplied by the factor $k$, accounting for the growth of the density of states $\rmd^{2}k\equiv k\rmd k\rmd \phi_{\bm k} $ with the absolute value of the wave vector.  The transfer rate Eq.~\eqref{eq:G2} is proportional to the integral of the product of the values of $k|\bm G_{DA,\bm k,\lambda}|^{2}$ in Fig.~\ref{fig:4Green}(a) and $\tau_{\bm k,
\lambda} $ in Fig.~\ref{fig:4Green}(b) over the Bloch wave vector $k$. Importantly, 
 due to the difference in the Green function values $|\bm G_{DA,\bm k,\lambda}|^{2}$, the out-of-plane transfer is mediated by the $p$-polarized  Bloch modes, while the in-plane transfer is controlled by the $s$-like modes. These modes have quite different lifetimes, that   are plotted in Fig.~\ref{fig:4Green}b for the height $h=20$~nm above the mirror (see also  Fig.~\ref{fig:2transfer}).  This explains the qualitative difference between the dependence of the in-plane transfer and the out-of-plane transfer on the distance from the mirror.
 %One can see, that for large wave vectors, when the coupling of the evanescent modes to the mirror is negligible, their lifetime is independent of the wave vector and is equal to $1/2\Gamma_{D}$, i.e. it is determined by the individual donor properties.
  In order to elucidate the effect  we will now analyze these two cases in more detail. (i) Out-of-plane transfer.
 When the array approaches the mirror, the lifetime of the $p$-type modes, determining the transfer, becomes shorter. The product
$W= \sum_{\bm k,\lambda}
\tau_{\bm k,\lambda}
|\bm G_{DA,\bm k,\lambda}|^{2}$, controlled by the $p$-type modes, is decreased. On the other hand, the total lifetime $T= \sum_{\bm k,\lambda}
\tau_{\bm k,\lambda}$ of the donor array is due to the long-living modes with $s$-type polarization, rather than the $p$ modes. The
lifetime of $s$-type modes is  equal to $1/2\Gamma_{D}$  for  $k\gtrsim 0.1\pi/a$  and   weakly sensitive to the mirror (red/dotted curve in Fig.~\ref{fig:4Green}b). As a result, $W$ decreases faster than the $T$ when the mirror is brought closer, and the transfer rate Eq.~\eqref{eq:G2} $\propto W/T$ is suppressed, in agreement with Fig.~\ref{fig:2transfer}(a). (ii) In-plane transfer. In this case the function $W$ is  determined by the long-living $s$-type eigenmode. Hence, the value of  $W$ is practically insensitive to the mirror because the lifetime is modified only for very small $k\lesssim 0.1\pi/a$, where the product of the squared Green function and the density of states is small. Hence, $W$ decreases slower than $T$, and the transfer rate $W/T$ is enhanced at smaller distances from the mirror,  in agreement with Fig.~\ref{fig:2transfer}(b).

%%%%%%%%%%%%%%%%%%%%%%%%%%%%%%%%%%%%%%%%%%%%%%%%%%%
\begin{figure}[t!]
\centering\includegraphics[width=0.45\textwidth]{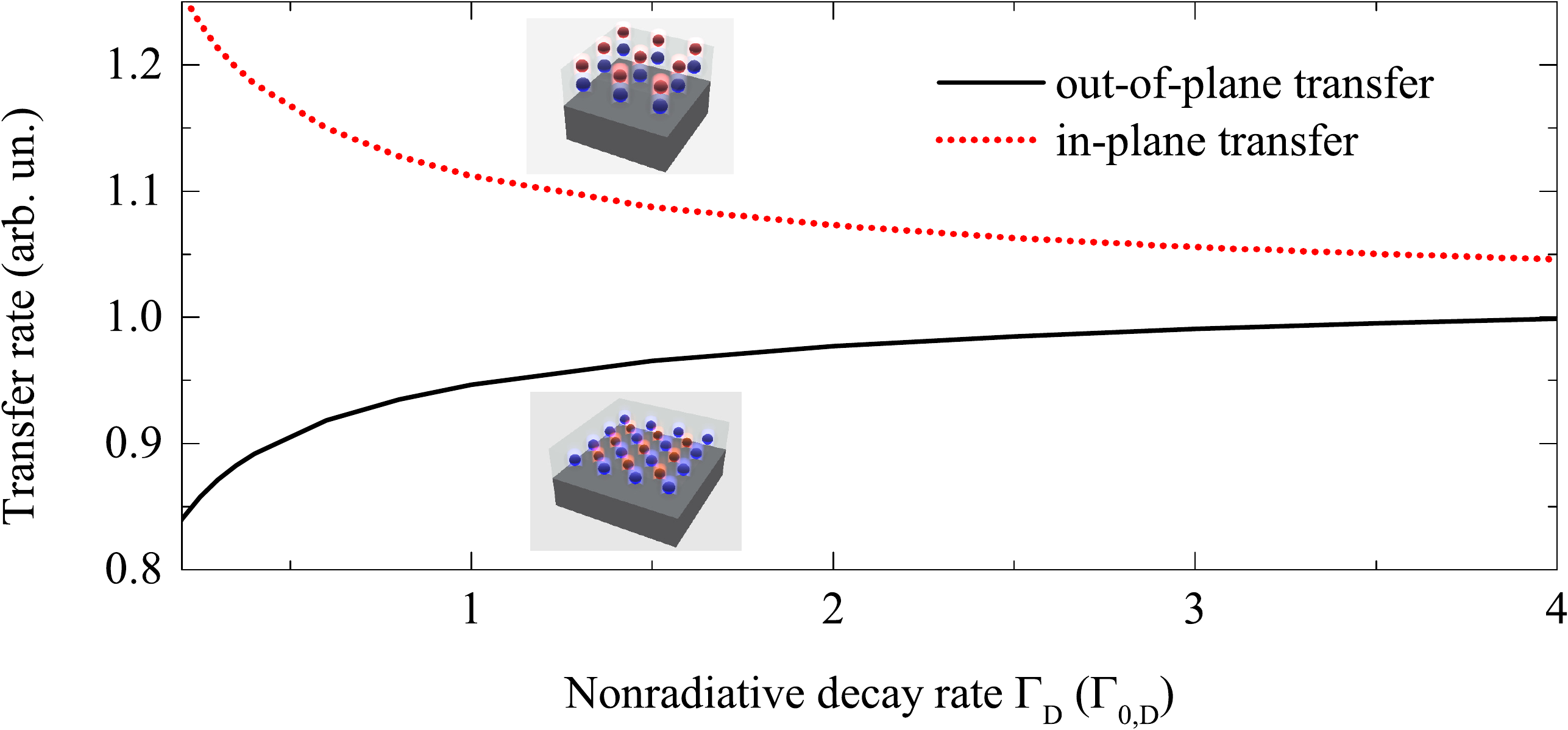}
\caption{
(Color online) Energy transfer rate as function of the nonradiative decay of the single donor $\Gamma$.
Black/solid and red/dotted curves correspond to the out-of-plane energy transfer and in-plane transfer, as indicated by the insets.
Other calculation parameters as the same as in Fig.~\ref{fig:2transfer}.
}\label{fig:5transfer}
\end{figure}
Finally, in Fig.~\ref{fig:5transfer} we analyze the dependence of the transfer rates on the nonradiative decay of the donors $\Gamma_{D}$. Clearly, when the nonradiative component of the decay rate increases, the decay rate of any given Bloch mode becomes less sensitive to the mirror presence. As a result, the total energy transfer rate returns to its bulk value, unaffected by the mirror. This is in agreement with the result of calculation in Fig.~\ref{fig:5transfer}: both for out-of-plane (red/dotted curve) and in-plane (black/solid) geometries the energy transfer tends to its bulk value for large $\Gamma_{D}$.

%%%%%%%%%%%%%%%%%%%%%%%%%%%%%%%%%%%%%%%%%%%%%%%%%%%

%%%%%%%%%%%%%%%%%%%%%%%%%%%%%%%%%%%%%%%%%%%%%%%%%%%
\section{Summary}\label{sec:summary}
To summarize, we have developed a theory of energy transfer between the planar arrays of donor and acceptor molecules, lying on top of a metallic mirror in the regime of weak,  incoherent and spatially uncorrelated pumping of the donors.
We have used a coupled dipole model with random source terms that has allowed us to obtain analytical results for the transfer rate. It has been demonstrated, that the transfer rate between the molecule arrays can be strongly modified by the mirror, in stark contrast to the transfer between individual molecules. The enhanced sensitivity of the   transfer to the presence of the mirror is due to the modification of the lifetimes of the collective Bloch modes of the donor array. Depending on the relative position between the donor and acceptor molecules, the transfer can be both suppressed (acceptors between the donors) or enhanced (acceptors above the donors).  While further study is needed to account for the possible effect of the strong coupling between the donors and the plasmons,\cite{Torma2014arXiv}  for inevitable effects of disorder in actual samples, and for the bulk film geometry, these our results might already provide the key to explain the suppression of the energy transfer in the plasmonic environment, recently observed experimentally in Ref.~\onlinecite{We2014}.

%%%%%%%%%%%%%%%%%%%%%%%%%%%%%%%%%%%%%%%%%%%%%%%%%%%

\section*{Acknowledgements}
I acknowledge useful discussions with 
M.A.~Noginov, A.V.~Rodina, E.L.~Ivchenko, E.E.~Narimanov,  T.V.~Shahbazyan, and P. Ginzburg.
This work has been funded by  the  Government of the Russian Federation (Grant No. 074-U01),
``Dynasty'' Foundation and RFBR.

%\bibliography{FRET}

\begin{thebibliography}{28}
\expandafter\ifx\csname natexlab\endcsname\relax\def\natexlab#1{#1}\fi
\expandafter\ifx\csname bibnamefont\endcsname\relax
  \def\bibnamefont#1{#1}\fi
\expandafter\ifx\csname bibfnamefont\endcsname\relax
  \def\bibfnamefont#1{#1}\fi
\expandafter\ifx\csname citenamefont\endcsname\relax
  \def\citenamefont#1{#1}\fi
\expandafter\ifx\csname url\endcsname\relax
  \def\url#1{\texttt{#1}}\fi
\expandafter\ifx\csname urlprefix\endcsname\relax\def\urlprefix{URL }\fi
\providecommand{\bibinfo}[2]{#2}
\providecommand{\eprint}[2][]{\url{#2}}

\bibitem[{\citenamefont{Agranovich et~al.}(2011)\citenamefont{Agranovich,
  Gartstein, and Litinskaya}}]{Agranovich2011}
\bibinfo{author}{\bibfnamefont{V.~M.} \bibnamefont{Agranovich}},
  \bibinfo{author}{\bibfnamefont{Y.~N.} \bibnamefont{Gartstein}},
  \bibnamefont{and}
  \bibinfo{author}{\bibfnamefont{M.}~\bibnamefont{Litinskaya}},
  \bibinfo{journal}{Chemical Reviews} \textbf{\bibinfo{volume}{111}},
  \bibinfo{pages}{5179} (\bibinfo{year}{2011}).

\bibitem[{\citenamefont{{Andreakou, P.} et~al.}(2013)\citenamefont{{Andreakou,
  P.}, {Brossard, M.}, {Li, C.}, {Lagoudakis, P. G.}, {Bernechea, M.}, and
  {Konstantatos, G.}}}]{Andreakou}
\bibinfo{author}{\bibnamefont{{Andreakou, P.}}},
  \bibinfo{author}{\bibnamefont{{Brossard, M.}}},
  \bibinfo{author}{\bibnamefont{{Li, C.}}},
  \bibinfo{author}{\bibnamefont{{Lagoudakis, P. G.}}},
  \bibinfo{author}{\bibnamefont{{Bernechea, M.}}}, \bibnamefont{and}
  \bibinfo{author}{\bibnamefont{{Konstantatos, G.}}}, \bibinfo{journal}{EPJ Web
  of Conferences} \textbf{\bibinfo{volume}{54}}, \bibinfo{pages}{01017}
  (\bibinfo{year}{2013}).

\bibitem[{\citenamefont{Rindermann et~al.}(2011)\citenamefont{Rindermann,
  Pozina, Monemar, Hultman, Amano, and Lagoudakis}}]{Rindermann}
\bibinfo{author}{\bibfnamefont{J.~J.} \bibnamefont{Rindermann}},
  \bibinfo{author}{\bibfnamefont{G.}~\bibnamefont{Pozina}},
  \bibinfo{author}{\bibfnamefont{B.}~\bibnamefont{Monemar}},
  \bibinfo{author}{\bibfnamefont{L.}~\bibnamefont{Hultman}},
  \bibinfo{author}{\bibfnamefont{H.}~\bibnamefont{Amano}}, \bibnamefont{and}
  \bibinfo{author}{\bibfnamefont{P.~G.} \bibnamefont{Lagoudakis}},
  \bibinfo{journal}{Phys. Rev. Lett.} \textbf{\bibinfo{volume}{107}},
  \bibinfo{pages}{236805} (\bibinfo{year}{2011}).

\bibitem[{\citenamefont{Fruhwirth et~al.}(2011)\citenamefont{Fruhwirth,
  Fernandes, Weitsman, Patel, Kelleher, Lawler, Brock, Poland, Matthews, K\'eri
  et~al.}}]{Fruhwirth}
\bibinfo{author}{\bibfnamefont{G.~O.} \bibnamefont{Fruhwirth}},
  \bibinfo{author}{\bibfnamefont{L.~P.} \bibnamefont{Fernandes}},
  \bibinfo{author}{\bibfnamefont{G.}~\bibnamefont{Weitsman}},
  \bibinfo{author}{\bibfnamefont{G.}~\bibnamefont{Patel}},
  \bibinfo{author}{\bibfnamefont{M.}~\bibnamefont{Kelleher}},
  \bibinfo{author}{\bibfnamefont{K.}~\bibnamefont{Lawler}},
  \bibinfo{author}{\bibfnamefont{A.}~\bibnamefont{Brock}},
  \bibinfo{author}{\bibfnamefont{S.~P.} \bibnamefont{Poland}},
  \bibinfo{author}{\bibfnamefont{D.~R.} \bibnamefont{Matthews}},
  \bibinfo{author}{\bibfnamefont{G.}~\bibnamefont{K\'eri}},
  \bibnamefont{et~al.}, \bibinfo{journal}{ChemPhysChem}
  \textbf{\bibinfo{volume}{12}}, \bibinfo{pages}{442} (\bibinfo{year}{2011}).

\bibitem[{\citenamefont{Galisteo-Lopez
  et~al.}(2014)\citenamefont{Galisteo-Lopez, Ibisate, and Lopez}}]{Lopez2014}
\bibinfo{author}{\bibfnamefont{J.~F.} \bibnamefont{Galisteo-Lopez}},
  \bibinfo{author}{\bibfnamefont{M.}~\bibnamefont{Ibisate}}, \bibnamefont{and}
  \bibinfo{author}{\bibfnamefont{C.}~\bibnamefont{Lopez}}, \bibinfo{journal}{J.
  Phys. Chem. C} \textbf{\bibinfo{volume}{118}}, \bibinfo{pages}{9665}
  (\bibinfo{year}{2014}).

\bibitem[{\citenamefont{Navarro-Urrios
  et~al.}(2009)\citenamefont{Navarro-Urrios, Pitanti, Daldosso, Gourbilleau,
  Rizk, Garrido, and Pavesi}}]{DaldossoPRB2009}
\bibinfo{author}{\bibfnamefont{D.}~\bibnamefont{Navarro-Urrios}},
  \bibinfo{author}{\bibfnamefont{A.}~\bibnamefont{Pitanti}},
  \bibinfo{author}{\bibfnamefont{N.}~\bibnamefont{Daldosso}},
  \bibinfo{author}{\bibfnamefont{F.}~\bibnamefont{Gourbilleau}},
  \bibinfo{author}{\bibfnamefont{R.}~\bibnamefont{Rizk}},
  \bibinfo{author}{\bibfnamefont{B.}~\bibnamefont{Garrido}}, \bibnamefont{and}
  \bibinfo{author}{\bibfnamefont{L.}~\bibnamefont{Pavesi}},
  \bibinfo{journal}{Phys. Rev. B} \textbf{\bibinfo{volume}{79}},
  \bibinfo{eid}{193312} (pages~\bibinfo{numpages}{4}) (\bibinfo{year}{2009}).

\bibitem[{\citenamefont{Izeddin et~al.}(2008)\citenamefont{Izeddin, Timmerman,
  Gregorkiewicz, Moskalenko, Prokofiev, Yassievich, and
  Fujii}}]{ProkofievTransferPRB2008}
\bibinfo{author}{\bibfnamefont{I.}~\bibnamefont{Izeddin}},
  \bibinfo{author}{\bibfnamefont{D.}~\bibnamefont{Timmerman}},
  \bibinfo{author}{\bibfnamefont{T.}~\bibnamefont{Gregorkiewicz}},
  \bibinfo{author}{\bibfnamefont{A.~S.} \bibnamefont{Moskalenko}},
  \bibinfo{author}{\bibfnamefont{A.~A.} \bibnamefont{Prokofiev}},
  \bibinfo{author}{\bibfnamefont{I.~N.} \bibnamefont{Yassievich}},
  \bibnamefont{and} \bibinfo{author}{\bibfnamefont{M.}~\bibnamefont{Fujii}},
  \bibinfo{journal}{Phys. Rev. B} \textbf{\bibinfo{volume}{78}},
  \bibinfo{eid}{035327} (\bibinfo{year}{2008}).

\bibitem[{\citenamefont{Agranovich and Galanin}(1982)}]{AgranovichGalanin}
\bibinfo{author}{\bibfnamefont{V.~M.} \bibnamefont{Agranovich}}
  \bibnamefont{and} \bibinfo{author}{\bibfnamefont{M.}~\bibnamefont{Galanin}},
  \emph{\bibinfo{title}{Electronic excitation energy transfer in condensed
  matter}} (\bibinfo{publisher}{North-Holland Pub. Co.},
  \bibinfo{address}{Amsterdam}, \bibinfo{year}{1982}).

\bibitem[{\citenamefont{Andrew and Barnes}(2000)}]{Andrew2000}
\bibinfo{author}{\bibfnamefont{P.}~\bibnamefont{Andrew}} \bibnamefont{and}
  \bibinfo{author}{\bibfnamefont{W.~L.} \bibnamefont{Barnes}},
  \bibinfo{journal}{Science} \textbf{\bibinfo{volume}{290}},
  \bibinfo{pages}{785} (\bibinfo{year}{2000}).

\bibitem[{\citenamefont{Nakamura et~al.}(2005)\citenamefont{Nakamura, Fujii,
  Imakita, and Hayashi}}]{Nakamura2005}
\bibinfo{author}{\bibfnamefont{T.}~\bibnamefont{Nakamura}},
  \bibinfo{author}{\bibfnamefont{M.}~\bibnamefont{Fujii}},
  \bibinfo{author}{\bibfnamefont{K.}~\bibnamefont{Imakita}}, \bibnamefont{and}
  \bibinfo{author}{\bibfnamefont{S.}~\bibnamefont{Hayashi}},
  \bibinfo{journal}{Phys. Rev. B} \textbf{\bibinfo{volume}{72}},
  \bibinfo{pages}{235412} (\bibinfo{year}{2005}).

\bibitem[{\citenamefont{{Rabouw} et~al.}(2014)\citenamefont{{Rabouw}, {den
  Hartog}, {Senden}, and {Meijerink}}}]{rabouw2014}
\bibinfo{author}{\bibfnamefont{F.~T.} \bibnamefont{{Rabouw}}},
  \bibinfo{author}{\bibfnamefont{S.~A.} \bibnamefont{{den Hartog}}},
  \bibinfo{author}{\bibfnamefont{T.}~\bibnamefont{{Senden}}}, \bibnamefont{and}
  \bibinfo{author}{\bibfnamefont{A.}~\bibnamefont{{Meijerink}}},
  \bibinfo{journal}{Nature Communications} \textbf{\bibinfo{volume}{5}},
  \bibinfo{eid}{3610} (\bibinfo{year}{2014}).

\bibitem[{\citenamefont{Blum et~al.}(2012)\citenamefont{Blum, Zijlstra,
  Lagendijk, Wubs, Mosk, Subramaniam, and Vos}}]{blum2012}
\bibinfo{author}{\bibfnamefont{C.}~\bibnamefont{Blum}},
  \bibinfo{author}{\bibfnamefont{N.}~\bibnamefont{Zijlstra}},
  \bibinfo{author}{\bibfnamefont{A.}~\bibnamefont{Lagendijk}},
  \bibinfo{author}{\bibfnamefont{M.}~\bibnamefont{Wubs}},
  \bibinfo{author}{\bibfnamefont{A.~P.} \bibnamefont{Mosk}},
  \bibinfo{author}{\bibfnamefont{V.}~\bibnamefont{Subramaniam}},
  \bibnamefont{and} \bibinfo{author}{\bibfnamefont{W.~L.} \bibnamefont{Vos}},
  \bibinfo{journal}{Phys. Rev. Lett.} \textbf{\bibinfo{volume}{109}},
  \bibinfo{pages}{203601} (\bibinfo{year}{2012}).

\bibitem[{\citenamefont{Tumkur et~al.}(2014)\citenamefont{Tumkur, Kitur,
  Bonner, Poddubny, Narimanov, and Noginov}}]{We2014}
\bibinfo{author}{\bibfnamefont{T.}~\bibnamefont{Tumkur}},
  \bibinfo{author}{\bibfnamefont{J.}~\bibnamefont{Kitur}},
  \bibinfo{author}{\bibfnamefont{C.}~\bibnamefont{Bonner}},
  \bibinfo{author}{\bibfnamefont{A.}~\bibnamefont{Poddubny}},
  \bibinfo{author}{\bibfnamefont{E.}~\bibnamefont{Narimanov}},
  \bibnamefont{and} \bibinfo{author}{\bibfnamefont{M.}~\bibnamefont{Noginov}},
  \bibinfo{journal}{Faraday Discuss.}  (\bibinfo{year}{2014}),
  \bibinfo{note}{accepted for publication, doi:10.1039/C4FD00184B}.

\bibitem[{\citenamefont{Juzeli\ifmmode~\bar{u}\else \={u}\fi{}nas and
  Andrews}(1994)}]{Andrews1994}
\bibinfo{author}{\bibfnamefont{G.}~\bibnamefont{Juzeli\ifmmode~\bar{u}\else
  \={u}\fi{}nas}} \bibnamefont{and} \bibinfo{author}{\bibfnamefont{D.~L.}
  \bibnamefont{Andrews}}, \bibinfo{journal}{Phys. Rev. B}
  \textbf{\bibinfo{volume}{49}}, \bibinfo{pages}{8751} (\bibinfo{year}{1994}).

\bibitem[{\citenamefont{Dung et~al.}(2002)\citenamefont{Dung, Kn\"oll, and
  Welsch}}]{Dung2002}
\bibinfo{author}{\bibfnamefont{H.~T.} \bibnamefont{Dung}},
  \bibinfo{author}{\bibfnamefont{L.}~\bibnamefont{Kn\"oll}}, \bibnamefont{and}
  \bibinfo{author}{\bibfnamefont{D.-G.} \bibnamefont{Welsch}},
  \bibinfo{journal}{Phys. Rev. A} \textbf{\bibinfo{volume}{65}},
  \bibinfo{pages}{043813} (\bibinfo{year}{2002}).

\bibitem[{\citenamefont{Andrews and Bradshaw}(2004)}]{Andrews2004}
\bibinfo{author}{\bibfnamefont{D.~L.} \bibnamefont{Andrews}} \bibnamefont{and}
  \bibinfo{author}{\bibfnamefont{D.~S.} \bibnamefont{Bradshaw}},
  \bibinfo{journal}{European Journal of Physics} \textbf{\bibinfo{volume}{25}},
  \bibinfo{pages}{845} (\bibinfo{year}{2004}).

\bibitem[{\citenamefont{Pustovit and Shahbazyan}(2011)}]{Shahbazyan2011}
\bibinfo{author}{\bibfnamefont{V.~N.} \bibnamefont{Pustovit}} \bibnamefont{and}
  \bibinfo{author}{\bibfnamefont{T.~V.} \bibnamefont{Shahbazyan}},
  \bibinfo{journal}{Phys. Rev. B} \textbf{\bibinfo{volume}{83}},
  \bibinfo{pages}{085427} (\bibinfo{year}{2011}).

\bibitem[{\citenamefont{Pustovit et~al.}(2013)\citenamefont{Pustovit, Urbas,
  and Shahbazyan}}]{Shahbazyan2013}
\bibinfo{author}{\bibfnamefont{V.~N.} \bibnamefont{Pustovit}},
  \bibinfo{author}{\bibfnamefont{A.~M.} \bibnamefont{Urbas}}, \bibnamefont{and}
  \bibinfo{author}{\bibfnamefont{T.~V.} \bibnamefont{Shahbazyan}},
  \bibinfo{journal}{Phys. Rev. B} \textbf{\bibinfo{volume}{88}},
  \bibinfo{pages}{245427} (\bibinfo{year}{2013}).

\bibitem[{\citenamefont{{Averkiev} et~al.}(2009)\citenamefont{{Averkiev},
  {Glazov}, and {Poddubnyi}}}]{JETP2009}
\bibinfo{author}{\bibfnamefont{N.~S.} \bibnamefont{{Averkiev}}},
  \bibinfo{author}{\bibfnamefont{M.~M.} \bibnamefont{{Glazov}}},
  \bibnamefont{and} \bibinfo{author}{\bibfnamefont{A.~N.}
  \bibnamefont{{Poddubnyi}}}, \bibinfo{journal}{JETP}
  \textbf{\bibinfo{volume}{108}}, \bibinfo{pages}{836} (\bibinfo{year}{2009}).

\bibitem[{\citenamefont{Toma\ifmmode~\check{s}\else
  \v{s}\fi{}}(1995)}]{Tomas1995}
\bibinfo{author}{\bibfnamefont{M.~S.} \bibnamefont{Toma\ifmmode~\check{s}\else
  \v{s}\fi{}}}, \bibinfo{journal}{Phys. Rev. A} \textbf{\bibinfo{volume}{51}},
  \bibinfo{pages}{2545} (\bibinfo{year}{1995}).

\bibitem[{\citenamefont{Novotny and Hecht}(2006)}]{Novotny2006}
\bibinfo{author}{\bibfnamefont{L.}~\bibnamefont{Novotny}} \bibnamefont{and}
  \bibinfo{author}{\bibfnamefont{B.}~\bibnamefont{Hecht}},
  \emph{\bibinfo{title}{{P}rinciples of {N}ano-{O}ptics}}
  (\bibinfo{publisher}{Cambridge University Press}, \bibinfo{address}{New
  York}, \bibinfo{year}{2006}).

\bibitem[{\citenamefont{Belov and Simovski}(2005)}]{Belov2005}
\bibinfo{author}{\bibfnamefont{P.~A.} \bibnamefont{Belov}} \bibnamefont{and}
  \bibinfo{author}{\bibfnamefont{C.~R.} \bibnamefont{Simovski}},
  \bibinfo{journal}{Phys. Rev. E} \textbf{\bibinfo{volume}{72}},
  \bibinfo{pages}{026615} (\bibinfo{year}{2005}).

\bibitem[{\citenamefont{{Kambe}}(1967)}]{Kambe1967}
\bibinfo{author}{\bibfnamefont{K.}~\bibnamefont{{Kambe}}},
  \bibinfo{journal}{Zeitschrift Naturforschung Teil A}
  \textbf{\bibinfo{volume}{22}}, \bibinfo{pages}{322} (\bibinfo{year}{1967}).

\bibitem[{\citenamefont{Vergeer et~al.}(2005)\citenamefont{Vergeer, Vlugt, Kox,
  den Hertog, van~der Eerden, and Meijerink}}]{Vergeer2005}
\bibinfo{author}{\bibfnamefont{P.}~\bibnamefont{Vergeer}},
  \bibinfo{author}{\bibfnamefont{T.~J.~H.} \bibnamefont{Vlugt}},
  \bibinfo{author}{\bibfnamefont{M.~H.~F.} \bibnamefont{Kox}},
  \bibinfo{author}{\bibfnamefont{M.~I.} \bibnamefont{den Hertog}},
  \bibinfo{author}{\bibfnamefont{J.~P. J.~M.} \bibnamefont{van~der Eerden}},
  \bibnamefont{and}
  \bibinfo{author}{\bibfnamefont{A.}~\bibnamefont{Meijerink}},
  \bibinfo{journal}{Phys. Rev. B} \textbf{\bibinfo{volume}{71}},
  \bibinfo{pages}{014119} (\bibinfo{year}{2005}).

\bibitem[{\citenamefont{Johnson and Christy }(1972)}]{JohnsonChristy}
\bibinfo{author}{\bibfnamefont{P.~B.} \bibnamefont{Johnson}} \bibnamefont{and}
  \bibinfo{author}{\bibfnamefont{R.~W.} \bibnamefont{Christy }},
  \bibinfo{journal}{Phys. Rev. B} \textbf{\bibinfo{volume}{6}},
  \bibinfo{pages}{4370} (\bibinfo{year}{1972}).

\bibitem[{\citenamefont{{Khitrova} and {Gibbs}}(2007)}]{Khitrova2007nat}
\bibinfo{author}{\bibfnamefont{G.}~\bibnamefont{{Khitrova}}} \bibnamefont{and}
  \bibinfo{author}{\bibfnamefont{H.~M.} \bibnamefont{{Gibbs}}},
  \bibinfo{journal}{Nat. Phys.} \textbf{\bibinfo{volume}{3}},
  \bibinfo{pages}{84} (\bibinfo{year}{2007}).

\bibitem[{\citenamefont{Ivchenko and Kavokin}(1992)}]{Kavokin1991}
\bibinfo{author}{\bibfnamefont{E.~L.} \bibnamefont{Ivchenko}} \bibnamefont{and}
  \bibinfo{author}{\bibfnamefont{A.~V.} \bibnamefont{Kavokin}},
  \bibinfo{journal}{Sov. Phys. Solid State} \textbf{\bibinfo{volume}{34}},
  \bibinfo{pages}{1815} (\bibinfo{year}{1992}).

\bibitem[{\citenamefont{{T{\"o}rm{\"a}} and {Barnes}}(2014)}]{Torma2014arXiv}
\bibinfo{author}{\bibfnamefont{P.}~\bibnamefont{{T{\"o}rm{\"a}}}}
  \bibnamefont{and} \bibinfo{author}{\bibfnamefont{W.~L.}
  \bibnamefont{{Barnes}}}, \bibinfo{journal}{ArXiv e-prints}
  (\bibinfo{year}{2014}), \eprint{1405.1661}.

\end{thebibliography}

\end{document}